\documentclass[journal]{IEEEtai}

\usepackage[colorlinks,urlcolor=blue,linkcolor=blue,citecolor=blue]{hyperref}

\usepackage{color,array}

\usepackage{graphicx}

\usepackage{natbib}
\usepackage{subfigure}
\usepackage{multirow}
\usepackage{pifont}
\usepackage{csquotes}
\usepackage{amsmath}
\usepackage{wrapfig}
\usepackage{amssymb}
\usepackage{booktabs}
\usepackage{algorithm}      % 提供 algorithm 环境
\usepackage{algorithmic}

\begin{document}

\title{ProLLaMA: A Protein Large Language Model for Multi-Task Protein Language Processing} 

\author{Liuzhenghao Lv, Zongying Lin, Hao Li, Yuyang Liu, Jiaxi Cui, Calvin Yu-Chian Chen, Li Yuan, Yonghong Tian, \IEEEmembership{Fellow, IEEE}
% \thanks{This paragraph of the first footnote will contain the date on which you submitted your paper for review. It will also contain support information, including sponsor and financial support acknowledgment. For example, ``This work was supported in part by the U.S. Department of Commerce under Grant BS123456.'' }
\thanks{This work was supported in part by the National Natural Science Foundation of China~(No. 62202014, No. 62332002, No. 62425101, No. 62088102).}
\thanks{Liuzhenghao Lv, Zongying Lin, Hao Li, Yuyang Liu, Calvin Yu-Chian Chen, Li Yuan and Yonghong Tian are with Peking University, China. Corresponding author: Li Yuan (yuanli-ece@pku.edu.cn), Yonghong Tian (yhtian@pku.edu.cn).}
\thanks{Hao Li, Li Yuan and Yonghong Tian are also with the Peng Cheng Laboratory, China.}
\thanks{Jiaxi Cui is with the Pandalla.AI, Beijing, China.}
\thanks{The project is available at \url{https://github.com/PKU-YuanGroup/ProLLaMA}.}
% \thanks{This paragraph will include the Associate Editor who handled your paper.}
}

\markboth{Journal of IEEE Transactions on Artificial Intelligence, Vol. 00, No. 0, Month 2020}
{Authors \MakeLowercase{\textit{Lv et al.}}: ProLLaMA: A Protein Large Language Model for Multi-Task Protein Language Processing}

\maketitle

\begin{abstract}
Recent advances in Protein Language Models (PLMs) have transformed protein engineering, yet unlike their counterparts in Natural Language Processing (NLP), current PLMs exhibit a fundamental limitation: they excel in either Protein Language Understanding (PLU) or Protein Language Generation (PLG), but rarely both. This fragmentation hinders progress in protein engineering. To bridge this gap, we introduce ProLLaMA, a multitask protein language model enhanced by the Evolutionary Protein Generation Framework (EPGF). We construct a comprehensive instruction dataset containing approximately 13 million samples with over 11,000 superfamily annotations to facilitate better modeling of sequence-function landscapes. We leverage a two-stage training approach to develop ProLLaMA, a multitask LLM with protein domain expertise. Our EPGF addresses the mismatch between statistic language modeling and biological constraints through three innovations: a multi-dimensional interpretable scorer, hierarchical efficient decoding, and a probabilistic-biophysical joint selection mechanism. Extensive experiments demonstrate that ProLLaMA excels in both unconditional and controllable protein generation tasks, achieving superior structural quality metrics compared to existing PLMs. Additionally, ProLLaMA demonstrates strong understanding capabilities with a 67.1\% exact match rate in superfamily prediction. EPGF significantly enhances the biological viability of generated sequences, as evidenced by improved biophysical scores (+4.3\%) and structural metrics (+14.5\%).
\end{abstract}

\begin{IEEEImpStatement}
Protein engineering is a rapidly evolving field that involves the design and modification of proteins to achieve desired functions, playing a critical role in diverse applications from medicine to materials science. Advances in this area have traditionally relied on labor-intensive experimental methods.
In this work, we demonstrate the potential of LLMs to advance protein engineering. Our model is capable of not only predicting protein properties, but also designing structurally plausible proteins with desired functions from scratch.
Moreover, EPGF enhances protein generation by integrating biophysical constraints, improving the biological viability of designed proteins and expanding the practical applications of AI-driven protein engineering.
\end{IEEEImpStatement}


\section{Introduction} \label{sec:introduction}
Large Language Models (LLMs), such as GPT-4 and LLaMA-2~\citep{achiam2023gpt4,touvron2023llama2}, have achieved outstanding performance in handling a wide range of Natural Language Processing (NLP) tasks~\citep{tamkin2021understanding,zhao2023survey,kocon2023chatgpt,huang2023chatgpt,zhong2023can,bang2023multitask}, including both Natural Language Generation (NLG) and Natural Language Understanding (NLU) tasks, in a generative manner. This surge in LLMs has extended their applications beyond traditional contexts, including their adoption in the challenging field of protein engineering~\citep{pmlr-v162-notin22a,strokach2022deep,ferruz2022controllable,fang2023mol,pei2024biot5+}.

Recent advances in Protein Language Models (PLMs) have opened new possibilities for protein engineering, leveraging large-scale protein sequence corpora to learn meaningful representations. However, existing PLMs exhibit a fundamental limitation: they lack multitasking capabilities, with most models excelling in either Protein Language Understanding (PLU)\citep{devlin2018bert,meier2021language,rives2021biological,brandes2022proteinbert} or Protein Language Generation (PLG)\citep{ferruz2022protgpt2,madani2023large,nijkamp2023progen2}, but rarely both. This stands in contrast to LLM in NLP. It creates a fragmented approach to protein engineering, where different tasks require different models.

Among these two tasks, PLG is generally more challenging than PLU. While understanding mainly requires extracting meaningful features from existing sequences, generation demands the production of novel, functional proteins that adhere to strict biophysical and evolutionary constraints~\cite {ferruz2022controllable,rives2021biological}. Current PLMs, despite their ability to generate statistically plausible sequences, often struggle to ensure biological viability, limiting their practical applications in protein engineering~\cite {ferruz2022protgpt2}.

We hypothesize that this limitation arises from the inherent mismatch between NLP decoding strategies such as nuclear sampling and beam search~\cite{freitag2017beam}, and the nature of biological sequences. Unlike natural language, where fluency and coherence are primary concerns, protein sequences must fold into stable three-dimensional structures and maintain biochemical functionality. NLP decoding strategy, widely adopted in PLMs, optimizes for sequence likelihood but lacks explicit biophysical constraints, leading to sequences that are statistically coherent yet functionally deficient~\cite{valentini2023promises}.

To address these challenges, we introduce ProLLaMA, a multitask protein language model that bridges the gap between understanding and generation, enhanced by our Evolutionary Protein Generation Framework (EPGF). EPGF improves PLMs by incorporating explicit biophysical guidance during inference, ensuring that generated sequences adhere to essential biological constraints. Specifically:

We construct a large-scale instruction dataset that contains approximately 13 million samples and encompasses both generation and understanding. We introduce a two-stage training framework to obtain ProLLaMA. In the first training stage, we leverage a pre-trained general LLM~(LLaMA-2-7B) to continually learn the protein language while maintaining the natural language knowledge. In the second stage, the model is further trained on the aforementioned instruction dataset. During inference, we use EPGF, a test-time computing framework, to enhance ProLLaMA's generation capabilities through three key innovations: (1) a Multi-dimensional Biophysical Scorer that evaluates sequences based on compositional biophysics, physicochemical properties, and functional characteristics; (2) a Hierarchical Efficient Decoding strategy that processes sequences at the segment level, aligning with natural protein folding patterns; and (3) a Probabilistic-Biophysical Joint Selection mechanism that dynamically balances statistical likelihood with biological viability.

\begin{figure*}[t] % 将宽度调整为0.4\textwidth
\centering
\includegraphics[width=0.95\textwidth]{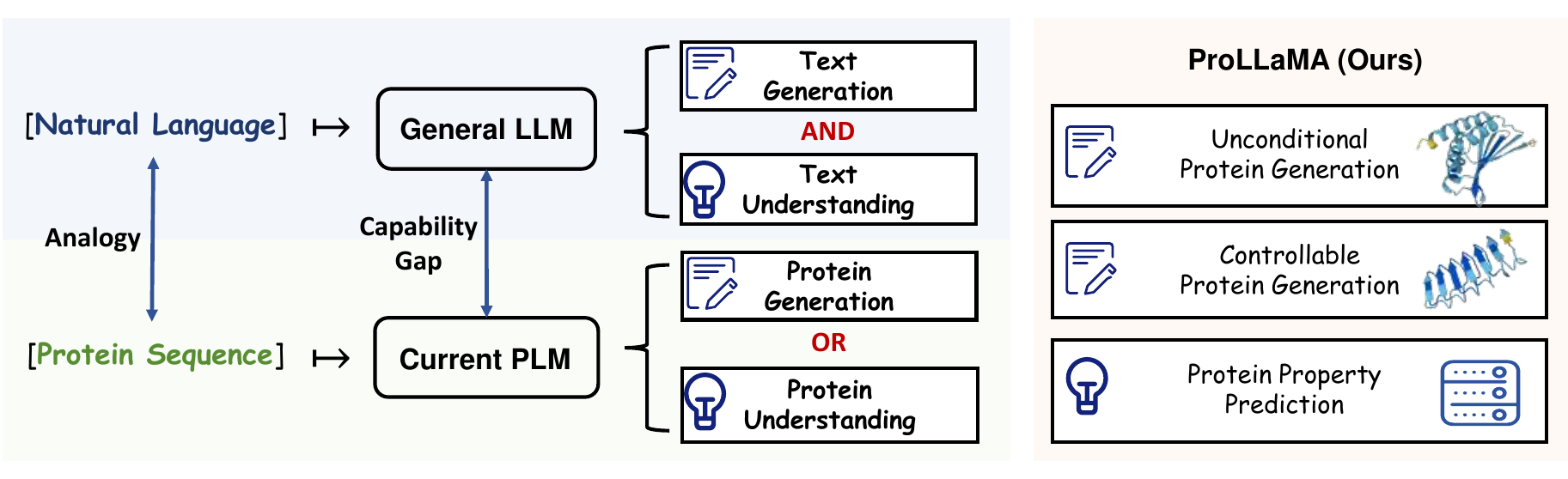} 
\vspace{-1.0\baselineskip}
\caption{\textbf{Left}: LLMs can handle both generation and understanding tasks, whereas PLMs cannot. This highlights the disparity in capabilities between the two. \textbf{Right}: Our ProLLaMA can handle generation tasks (unconditional protein generation, controllable protein generation) and understanding tasks (protein superfamily prediction), surpassing current PLMs.} 
% \label{figure:intro} 
\end{figure*}

Through extensive experiments, we demonstrate the multi-task capabilities of ProLLaMA and the effectiveness of EPGF. In unconditional protein generation, ProLLaMA outperforms current PLMs on common metrics such as pLDDT and TM-score. In controllable protein generation, ProLLaMA generates novel proteins from scratch with desired functionalities, such as the SAM-MT superfamily, based on user-provided textual descriptions. For protein superfamily prediction, ProLLaMA achieves a 67.1\% exact match rate on the test dataset and obtains an F1-score above 0.9 in many specific categories. Furthermore, EPGF significantly enhances the biological viability of generated sequences, as evidenced by improved biophysical scores~(+4.3\%) and structural metrics~(+14.5\%).

In summary, the contributions of our research are as follows:

$\bullet$~We construct an instruction dataset that contains 13 million samples and more than 11,000 kinds of superfamily annotations, which facilitates better modeling of sequence-function landscapes and enables multitask learning in the protein domain.

$\bullet$~We propose ProLLaMA, a multitask protein language model that bridges the gap between protein generation and understanding.

$\bullet$~We propose the Evolutionary Protein Generation Framework (EPGF), which ensures that generated protein sequences are not only statistically coherent but also biologically viable, addressing a critical limitation in current PLMs.

$\bullet$~Through extensive experiments, we demonstrate that ProLLaMA, enhanced by EPGF, achieves state-of-the-art results in protein generation tasks while excelling in protein understanding tasks.

\begin{figure*}[t]
\centering     %%% not \center
\includegraphics[width=0.8\textwidth]{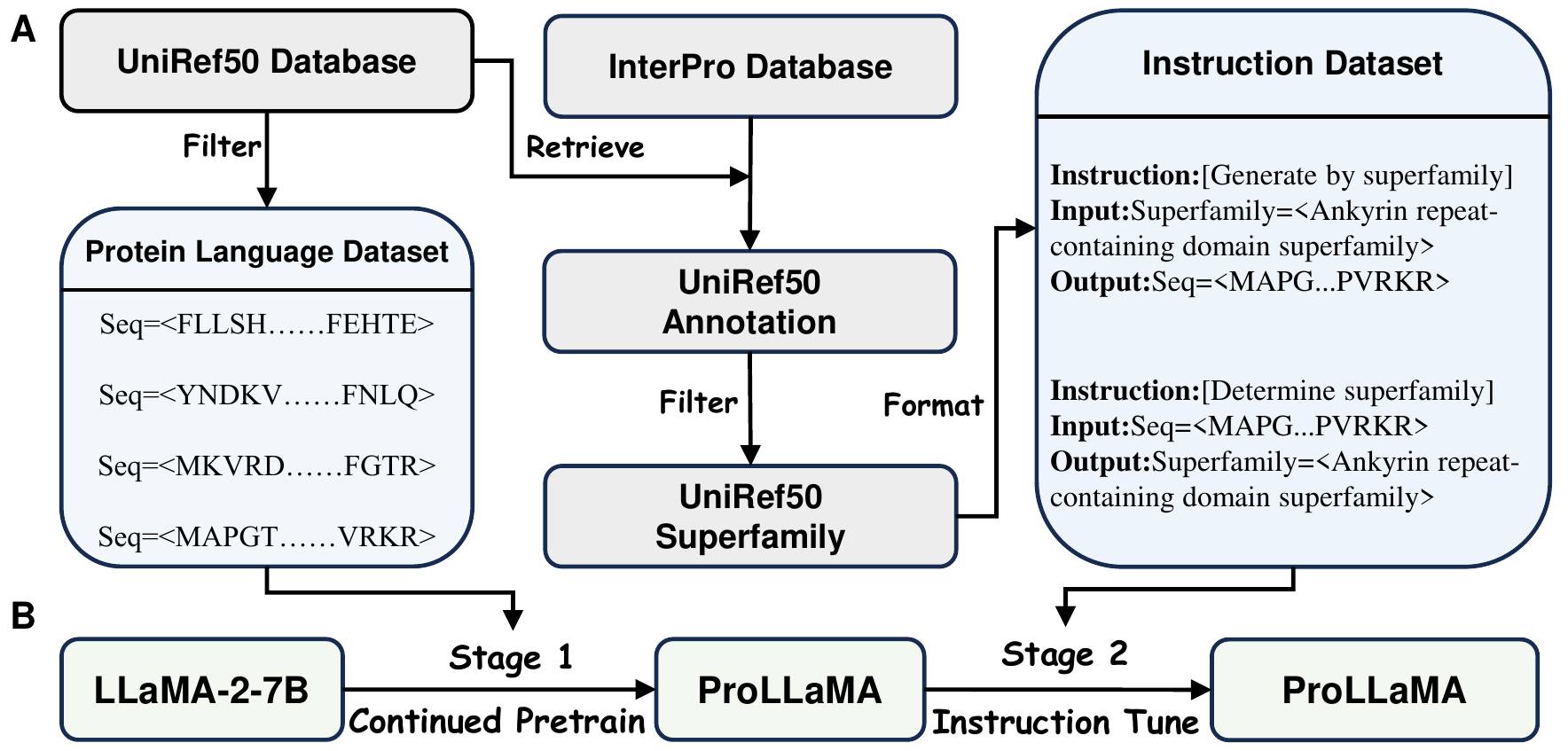} 
\caption{(A) \textbf{Overview of the dataset construction.} The protein language dataset contains 53 million samples, which is used for training in Stage 1. The instruction dataset contains 13 million instances with 11,268 unique superfamily annotations, which is used for training in Stage 2. (B) \textbf{Overview of the training framework.} Stage 1: The pre-trained LLaMA-2 learns the protein language, resulting in ProLLaMA. Stage 2: ProLLaMA learns to perform multiple tasks by instruction tuning.}\label{fig:pipeline}
\end{figure*}

\begin{figure*}[t]
\centering    
\includegraphics[width=0.95\textwidth]{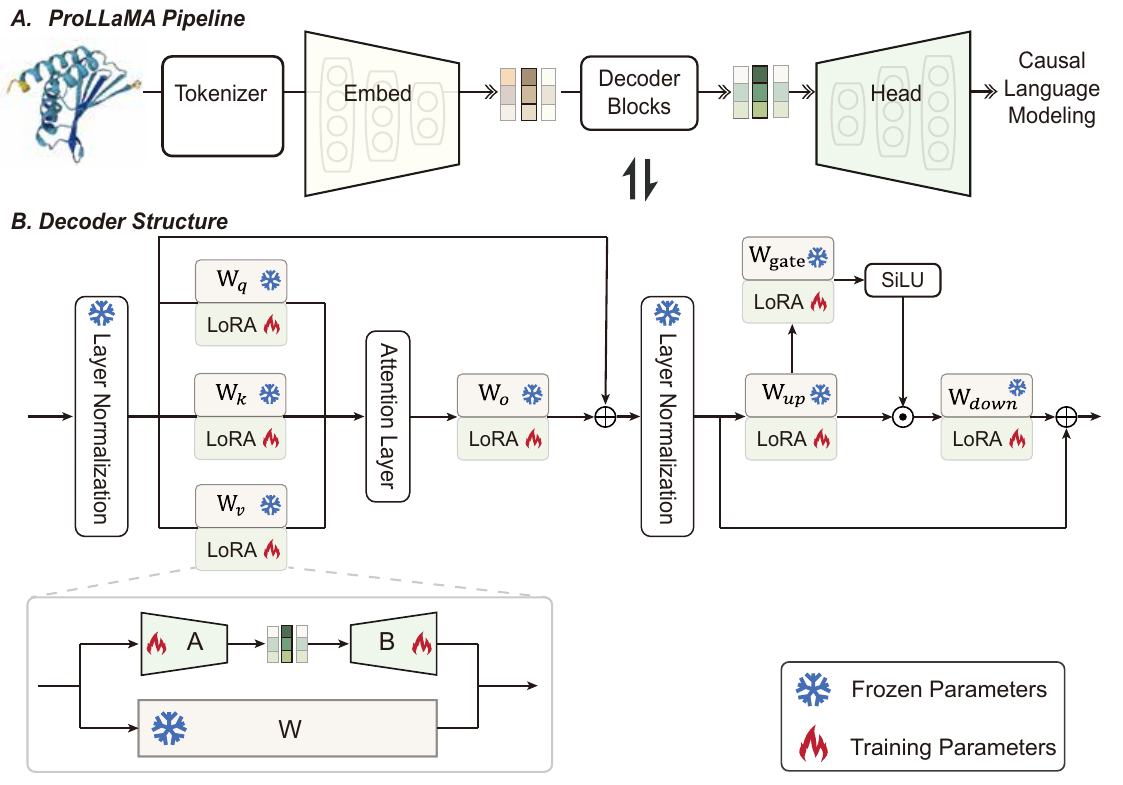} 
\vspace{-0.7\baselineskip}
\caption{\textbf{The overview of the ProLLaMA model.} We add LoRA adapters to certain weights. We freeze original parameters, focusing solely on training LoRA adapters (\textit{Embed} and \textit{Head} are also involved in the first training stage).} 
\label{fig:model}
\end{figure*}

\begin{figure*}
    \centering
    \includegraphics[width=0.9\linewidth]{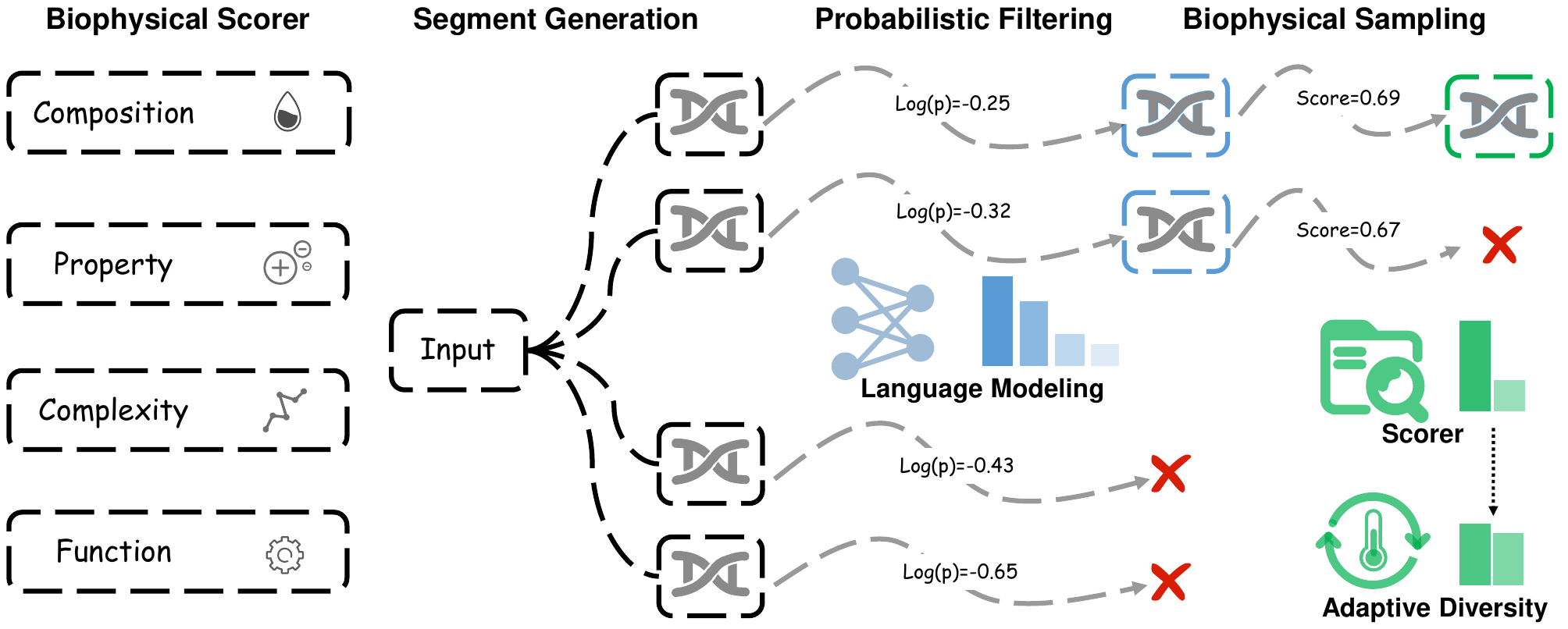}
    \vspace{-0.7\baselineskip}
    \caption{\textbf{The overview of EPGF.} EPGF has three key components: (1) a multi-dimensional biophysical scorer; (2) a hierarchical efficient decoding strategy which generates protein candidates at segment-level; (3) probabilistic-biophysical joint selection with adaptive diversity control, which selects the superior candidate for the next round of generation.}
    \label{fig:epgf_framework}
\end{figure*}

\section{Methods} \label{sec:methods}
In Section.~\ref{sec:dataset} and Fig.~\ref{fig:pipeline}A, we show how to construct the protein language dataset and the instruction dataset. 
In Section.~\ref{sec:training-framework} and Fig.~\ref{fig:pipeline}B., we show how to develop ProLLaMA using our training framework.
In Section.~\ref{sec:EPGF}, Fig.~\ref{fig:epgf_framework} and Algorithm.~\ref{alg:epgf}, we demonstrate Evolutionary Protein Generation Framework~(EPGF).

\subsection{Dataset Construction} \label{sec:dataset}

\textbf{The protein language dataset} is utilized in the first training stage to enable LLaMA-2-7B to grasp the language of proteins. Specifically, the dataset is sourced from UniRef50\_2023\_03~\citep{suzek2015uniref} on the UniProt website. We eliminate the descriptive parts of UniRef50, retaining only the pure protein sequences.  Furthermore, We filter UniRef50 to ensure that the protein sequences consisted only of the 20 standard amino acids. We also
retain sequences with a length of less than 512, aligning with ProGen~\citep{madani2023large}. Given that the lengths of protein sequences follow a long-tail distribution, the sequences that are deleted constitute only a small portion of the total dataset. To preprocess the protein sequences, we employ a specific prefix and a suffix. This standardized format aids LLaMA-2 in distinguishing the new protein language from its existing natural language knowledge, thus reducing confusion. The original Uniref50 contains 60,952,894 sequences, while after the above series of processing, our dataset comprises 52,807,283 protein sequences, with 90\% for training and 10\% reserved for testing.

\textbf{The instruction dataset} is utilized in the second training stage to enable ProLLaMA to perform various tasks. We first obtain the protein2ipr database from InterPro~\citep{paysan2023interpro}, which includes all proteins from UniProtKB along with their corresponding InterPro annotation information. Subsequently, we iterate through each protein's rep\_id in UniRef50 to retrieve the corresponding annotation information from protein2ipr. This retrieval process is implemented using a distributed Redis database, and only proteins with lengths less than 256 participate in the retrieval to enhance efficiency. We utilize regular expressions to filter out superfamily annotation from the whole annotation. In the end, we obtain 6,350,106 data instances, each of which contains one protein sequence and its superfamily annotation. And the number of unique superfamily annotations is 11,268.

Then, we process the obtained data into a multi-task instruction dataset following the Alpaca format~\citep{alpaca}, where each instance comprises three parts: instruction, input, and output. The instruction specifies the task type. We design two tasks: generating proteins based on superfamily and determining the superfamily of the given protein. For the former task, the input is the superfamily annotation, and the output is the expected protein. The latter task is the opposite. In the end, the instruction dataset comprises 12,700,212 (6,350,106 * 2) instances, with 90\% for training and the rest reserved for testing.

\subsection{Training Framework for ProLLaMA} \label{sec:training-framework}

We propose a parameter-efficient training framework to transform general LLMs into PLMs capable of handling multiple tasks. Our framework leverages Low-Rank Adaptation (LoRA)~\citep{hu2021lora}, which injects protein-related knowledge into LLaMA-2 while preserving its natural language capabilities.

LoRA is a parameter-efficient technique that freezes the original LLM parameters and introduces trainable low-rank adapters. Theoretically, fine-tuning can be conceptualized as finding the parameter change $\Delta W = W - W_0$, where $W_0$ and $W$ represent the original and fine-tuned parameters, respectively. Assuming $\Delta W$ has a low rank $r$~\cite{aghajanyan2021intrinsic}, it can be decomposed as $\Delta W = AB$, yielding:
\begin{equation}
\textit{W} = \textit{W}_0 + \textit{AB} \label{eq:lora}
\end{equation}
where $W, W_0 \in \mathbb{R}^{d \times h}$, $A \in \mathbb{R}^{d \times r}$, and $B \in \mathbb{R}^{r \times h}$. This reduces trainable parameters from $dh$ to $r(d+h)$, with $r \ll d$ and $r \ll h$.

Our training framework consists of two key stages:

\textbf{Stage 1: Learning Protein Language.} We apply LoRA adapters to specific weights in each LLaMA-2 decoder block, including $W_q$, $W_k$, $W_v$, $W_o$, $W_{up}$, $W_{gate}$, and $W_{down}$. Due to significant differences between protein and natural languages, we use a relatively high rank for LoRA to prevent underfitting. We also train the embedding and output layers since tokens may have different meanings in protein sequences versus natural language. Through causal language modeling on protein sequences, we train only about 8\% of the parameters, which substantially reduces computational costs while preserving natural language abilities.

\textbf{Stage 2: Instruction Tuning.} To enable multi-task capabilities, we perform instruction tuning on the model from Stage 1. The training objective is:
\begin{equation}
\mathcal{L}(\Theta) = \mathbb{E}_{\boldsymbol{x,u} \sim \mathcal{D}}\left[ - \sum_i \log p(x_i | \boldsymbol{u},x_0, x_1, \ldots, x_{i-1} ; \Theta) \right] \label{eq:sft}
\end{equation}
where $\boldsymbol{u}$ denotes the instruction (including input) and $\boldsymbol{x} = \{x_0, x_1, \ldots, x_{n-1}\}$ represents the expected output. During this stage, we exclusively train LoRA at a lower rank than in Stage~1.

Our framework is flexible and extensible, allowing ProLLaMA to be easily adapted to additional tasks. Researchers can customize instruction datasets and perform further instruction tuning on ProLLaMA with minimal training resources. In our experiments, we demonstrate this extensibility by successfully adapting ProLLaMA to protein solubility prediction through additional instruction tuning.

The framework enables ProLLaMA to understand both protein language and natural language, follow instructions, and perform multiple protein-related tasks efficiently.

\begin{algorithm}[t] % 使用 [H] 强制算法在当前位置
\caption{Evolutionary Protein Generation Framework (EPGF)}
\label{alg:epgf}
\centering
\begin{algorithmic}[1]
\REQUIRE Language modeling distribution $p$, segment length $L$, diversity controller $\tau$, biophysical scorer $\mathcal{B}$, number of candidates $N$, minimum acceptable score $e$
\ENSURE Generated protein sequence $S$.

\STATE Initialize sequence $S_0$ with start token.
\STATE Initialize temperature $\tau_0$ and decay rate $\gamma$.

\WHILE{$S$ not finished}
    \STATE Sample $N$ candidate segments:
    
    $\{C_1, C_2, \dots, C_N\} \sim p,  |C_i| \leq N$
    
    \FOR{each candidate segment $C_j$}
        \STATE Log-probability $P(C_j) = \sum_{i=1}^L \log p(c_i \mid c_{<i})$.
    \ENDFOR

    \STATE Retain top $K = \lceil N/2 \rceil$ candidates based on $P(C_j)$.

    \FOR{each retained candidate $C_j$}
        \STATE Compute biophysical score $\mathcal{B}(C_j)$.
        \STATE Ensure $\mathcal{B}(C_j) \geq e$
    \ENDFOR
    \STATE Compute selection probability:
    \[
    P_{\text{select}}(C_j) = \frac{\exp(\mathcal{B}(C_j) / \tau_t)}{\sum_{k=1}^K \exp(\mathcal{B}(C_k) / \tau_t)}.
    \]
    \STATE Sample candidate: $C^* \sim P_{\text{select}}$.

    \STATE Update Sequence: $S=S+C^*$.

    \STATE Update Diversity Controller: $\tau_{} = \max(\tau_{\text{final}}, \tau \cdot \gamma)$.
\ENDWHILE

\RETURN Final sequence $S$.
\end{algorithmic}
\end{algorithm}

\subsection{Evolutionary Protein Generation Framework} \label{sec:EPGF}

We propose the \textbf{Evolutionary Protein Generation Framework (EPGF)}, a novel inference framework designed to enhance the sequence generation capabilities of PLMs by explicitly incorporating biological constraints. EPGF bridges the gap between statistical language modeling and functional protein design through three key innovations:

\begin{itemize}
    \item \textbf{Multi-dimensional Biophysical Scorer}: A biologically interpretable scoring system that evaluates protein sequences based on compositional biophysics, physicochemical properties, sequence complexity, and functional characteristics.
    \item \textbf{Hierarchical Efficient Decoding}: A structure-aware generation paradigm that processes sequences at the segment level (approximately 30 amino acids), aligning with natural protein folding patterns and significantly improving evaluation efficiency.
    \item \textbf{Probabilistic-Biophysical Joint Selection with Adaptive Diversity Control}: A unified mechanism that combines statistical likelihood with biological viability assessment, dynamically balancing exploration and exploitation during sequence generation through a simulated annealing-inspired approach.
\end{itemize}

We demonstrate the effectiveness of EPGF and its individual components through comprehensive experiments, including ablation studies, as presented in the experimental section and Supplementary Material.

\subsubsection{Multi-dimensional Biophysical Scorer}

%我们的提出的scorer旨从以下4个方面对蛋白质序列进行全方位的评估，并通过将关键metric进行加权平均给出综合性得分：
Our proposed scorer aims to comprehensively evaluate protein sequences from the following four aspects and provides an overall score by weighting and averaging key metrics:
\begin{equation}
\mathcal{B}(S) = \frac{1}{n}\sum_{i=1}^{n} \text{Metric}_i(S),
\end{equation}
where $S$ represents a protein sequence, and $\text{Metric}_i$ represents the $i$-th metric drawn from four major categories:
\begin{itemize}
\item \textbf{Compositional Biophysics}: Evaluates amino acid distribution, diversity, and prevalence of rare amino acids to ensure natural sequence composition. Research~\cite{gromiha1999aa} shows proteins with abnormal amino acid distributions typically exhibit reduced stability and functionality. Rare amino acids like cysteine and tryptophan serve critical roles but their improper distribution can compromise structure.

\item \textbf{Physicochemical Properties}: Assesses hydrophobicity, charge balance, and sequence stability, which are crucial for proper protein folding~\cite{kellis1988pp}. Hydrophobicity patterns drive tertiary structure formation with hydrophobic residues clustering in the protein core. Charge distribution affects solubility and stability, with imbalanced charges linked to aggregation.

\item \textbf{Sequence Complexity}: Measures sequence entropy, repetitive patterns, and local complexity to avoid unnatural homopolymers or overly repetitive sequences. Low-complexity regions rarely occur in functional globular proteins and often indicate intrinsically disordered regions or pathological aggregation~\cite{das2014sc}. Sequences with abnormally low complexity frequently fail to form stable structures.

\item \textbf{Functional Characteristics}: Examines secondary structure propensities and functional motifs associated with specific protein superfamilies. Well-balanced secondary structure elements demonstrate better stability and functionality. Conserved functional motifs represent evolutionarily optimized patterns essential for specific biochemical functions and activity~\cite{koehl1999fc}.
\end{itemize}

The specific calculation formula is shown in Supplementary Material. The proposed scorer enables quantitative evaluation of both complete and partial protein sequences, providing explicit insights into their biophysical properties during the design process. By assessing key metrics across multiple dimensions, it identifies potential issues such as unnatural amino acid distributions, improper physicochemical profiles, or structural anomalies, guiding the generation process of PLMs.

% This multi-faceted evaluation ensures that sequences generated by ProLLaMA maintain biological plausibility across various dimensions simultaneously. Unlike traditional approaches that focus solely on statistical likelihood, the scorer explicitly incorporates domain knowledge about protein structure and function.
%该scorer借助BioPython实现，以蛋白质序列作为输入而非昂贵的3D结构数据，因此能实现快速的高通量的评估（在5.5s完成对10000条序列的评估）
% The scorer is implemented using BioPython, taking protein sequences as input instead of costly 3D structural data, enabling rapid high-throughput evaluation (assessing 10,000 sequences in just 5.5 seconds). The interpretability of the scorer lies in its biologically grounded metrics, each reflecting well-established biochemical principles. This allows users to intuitively understand how scores are assigned and diagnose specific sequence deficiencies. The composite BioScore maintains transparency through the independent contribution of each metric, enabling complete score traceability.

\begin{figure*}[t]
  \centering
  \includegraphics[width=1\textwidth]{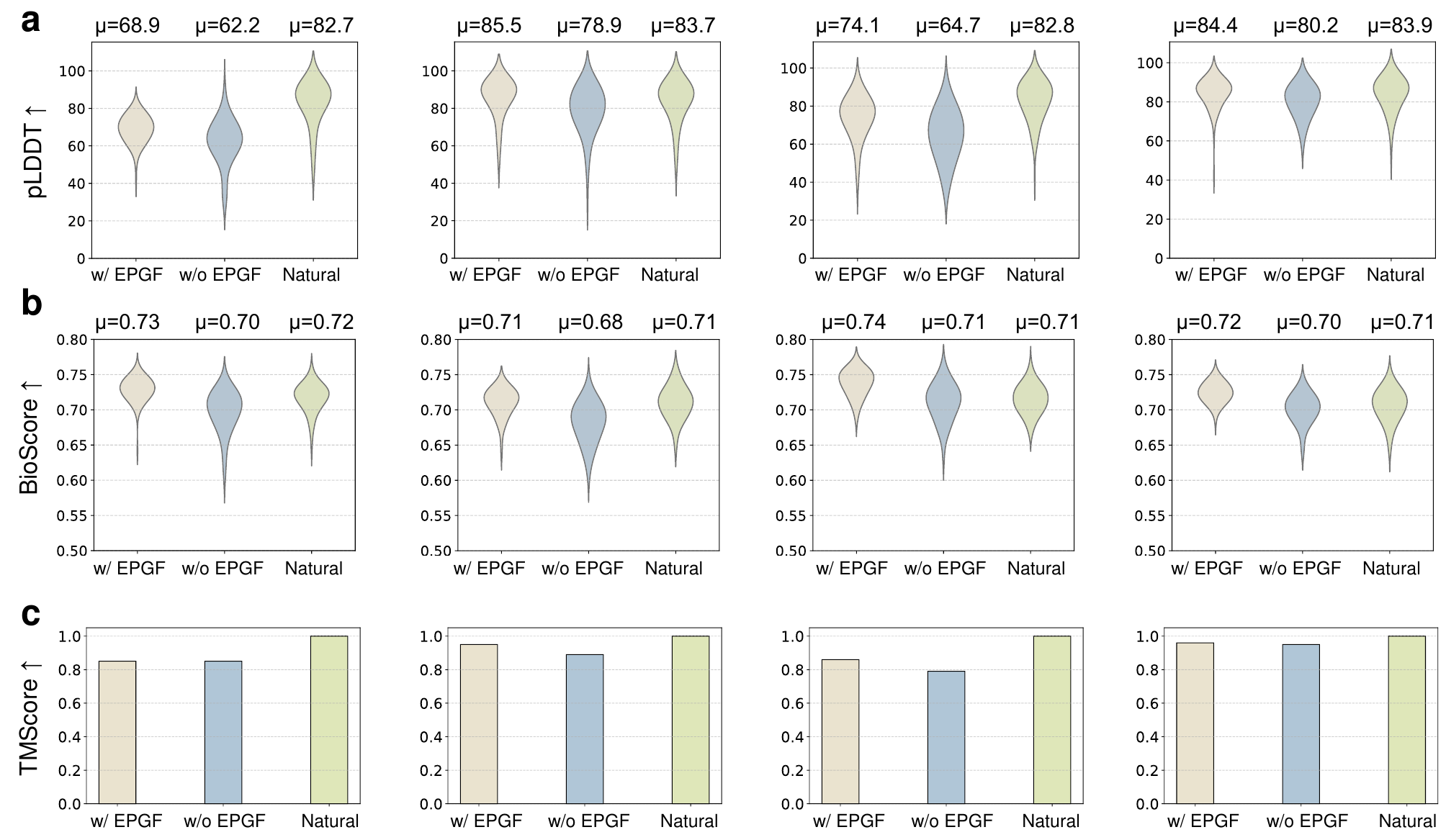}
  \vspace{-1.5\baselineskip}
  \caption{\textbf{ProLLaMA generates better protein sequences with EPGF.}  We visualize the (a) pLDDT (b) BioScore (c) TM-Score values of proteins belonging to four superfamiles (in order: SAM-MT, TPHD, Trx, CheY). $w/$: ProLLaMA with EPGF; $w/o$: ProLLaMA alone; $Natural$: Natural proteins as reference; $\mu$: the average value; $BioScore$: the biophysical score calculated by our scorer. EPGF improves the performance of ProLLaMA and even makes the generated proteins approach or even surpass the natural proteins on pLDDT.}
  \label{fig:epgf}
\end{figure*}

\subsubsection{Hierarchical Efficient Decoding}
%如果仅仅对完整的序列进行评估以选出优胜者，那么很可能掉入xxx陷阱。受到Beam Search的启发，我们在序列的生成过程中定期进行评估，以尽可能选出真正的优胜者。考虑到生物序列与自然语言的显著不同，我们提出了更加高效且生物合理的Segment-based Decoding。具体来说，不同于传统Beam Search中以token为单位的生成与评估，我们的EPGF以30个左右的氨基酸组成的segment为单位。This length corresponds to typical secondary structure elements like alpha-helices or beta-strands, which form the building blocks of protein domains. The biological motivation for this approach stems from the observation that proteins fold into discrete structural units, and generation should respect these natural boundaries. 
Traditional methods generate sequences token-by-token, which is inherently inefficient during evaluation, as the computational cost scales with residue numbers. It also poses challenges for meaningful biological evaluation, as assessing individual tokens in isolation lacks biological context and significance. To address these limitations, we propose a segment-level decoding strategy within EPGF. By evaluating sequences at the segment level, we not only significantly improve computational efficiency but also ensure the evaluation aligns with biologically relevant units, such as structural motifs and functional domains.

%Let $S = \{s_1, s_2, \dots, s_n\}$ denote a protein sequence of length $n$, where $s_i \in \mathcal{A}$ and $\mathcal{A}$ is the set of 20 canonical amino acids. 
Although sequence generation within each segment proceeds in a token-by-token manner, the principal innovation of our method resides in the evaluation paradigm. Instead of assessing individual tokens, we evaluate contiguous sequence fragments of length $L=20$ tokens (approximately 30 amino acids), which aligns with the characteristic size of typical protein structural motifs. The candidate segments $\{C_1, C_2, \dots, C_N\}$, where each $C_j$ consists of $L$ tokens $\{c_1, c_2, \dots, c_L\}$, are subsequently processed by our \textit{Probabilistic-Biophysical Joint Selection} mechanism (described below). This hierarchical efficient decoding strategy maintains the model's ability to generate high-quality sequences while reducing computational overhead.

\begin{table*}[t]
\centering
\footnotesize
\renewcommand{\arraystretch}{1.35}  % 控制行高
\setlength{\tabcolsep}{2.2mm}        % 控制列间距
{
\caption{\textbf{Comparison of proteins generated by different models.} Our ProLLaMA achieves the best performance on pLDDT, TM-score, and RMSD metrics, and is second-best in SC-Perp, demonstrating ProLLaMA excels in de novo protein design. AE: Auto-Encoder. AR: Auto-Regressive.} \label{table:prollama-compare}
\scalebox{1}{
\begin{tabular}{p{50pt}|c|cc|c@{\hspace{1mm}}c|c@{\hspace{1mm}}c}
    \toprule
    \multirow{2}{*}{\textbf{Type}} &\multirow{2}{*}{\textbf{Method}}&\multirow{2}{*}{\textbf{pLDDT$\uparrow$}}&\multirow{2}{*}{\textbf{SC-Perp$\downarrow$}}& \multicolumn{2}{c|}{\textbf{AFDB}} & \multicolumn{2}{c}{\textbf{PDB}}  \\
    \cmidrule(rl){5-6}\cmidrule(rl){7-8}
    &&&&\textbf{TM-score$\uparrow$} & \textbf{RMSD$\downarrow$}  & \textbf{TM-score$\uparrow$} & \textbf{RMSD$\downarrow$}  \\
    \midrule
    \multirow{2}{*}{CNN} & CARP~\citep{alamdari2023protein} & 34.40±14.43 & 4.05±0.52 & 0.28&19.38 & 0.38&8.95\\
    & LRAR~\citep{alamdari2023protein} & 49.13±15.50 & 3.59±0.54 &0.40&14.47& 0.43&9.47\\
    \midrule
    \multirow{2}{*}{PLM (AE)} & ESM-1b~\citep{rives2021biological}&59.57±15.36 &3.47±0.68& 0.34 & 20.88 & 0.44&8.59 \\
    & ESM-2~\citep{lin2023evolutionary}& 51.16±15.52 & 3.58±0.69& 0.20 & 35.70 & 0.41&9.57\\
    \midrule
    Diffusion & EvoDiff~\citep{alamdari2023protein} & 44.29±14.51 & 3.71±0.52& 0.32&21.02 & 0.41&10.11\\
    \midrule
    \multirow{3}{*}{PLM (AR)} & ProtGPT2~\citep{ferruz2022protgpt2} & 56.32±16.05&3.27±0.59 & 0.44 & 12.60 &  0.43&9.19 \\
    & ProGen2~\citep{nijkamp2023progen2} & 61.07±18.45&\textbf{2.90±0.71} & 0.43 & 15.52 & 0.44&11.02\\
    & \textbf{ProLLaMA}~(ours) & \textbf{66.49±12.61}& 3.10±0.65 & \textbf{0.49} & \textbf{9.50} & \textbf{0.48}&\textbf{7.63}\\
 \bottomrule
\end{tabular}
}
}
\end{table*}

\subsubsection{Probabilistic-Biophysical Joint Selection with Adaptive Diversity Control}

We propose a joint selection strategy that establishes an equilibrium between the statistical likelihood derived from language models and the biochemical validity determined by our specialized scorer. Models that optimize solely for probabilities often generate sequences that are statistically plausible but biologically aberrant. Conversely, methods prioritizing biochemical constraints may produce sequences lacking the statistical signatures characteristic of natural proteins. To address this limitation, the candidate segments are evaluated using a two-stage \textit{Probabilistic-Biophysical Joint Selection} mechanism, enhanced by an \textit{Adaptive Diversity Control} strategy to maintain sequence diversity.

\textbf{Probabilistic Filtering.} In the first stage, we evaluate the statistical coherence of the candidate segments. For each candidate segment  $C_j$, we compute its log-probability under the PLM:
\begin{equation}
    P(C_j) = \sum_{i=1}^L \log p(c_i \mid c_{<i}),
\end{equation}
where \( p(c_i \mid c_{<i}) \) is the conditional probability of token \( c_i \) given the preceding tokens \( c_{<i} \). The top \( K \) candidates with the highest \( P(C_j) \) are retained for further evaluation, where \( K \) is defined as $K = \left\lceil \frac{N}{2} \right\rceil,$
and \( N \) is the total number of candidate segments. This step ensures that the generated sequences are statistically fluent within the PLM's learned distribution.

\textbf{Biophysical Sampling with Adaptive Diversity Control}. The retained candidates are then evaluated using the proposed \textit{Multi-dimensional Biophysical Scorer}, with the selection process dynamically regulated by our \textit{Adaptive Diversity Control} strategy. This integration addresses the potential limitation of excessive filtering, which might lead to reduced sequence diversity.

For each candidate segment \( C_j \), we compute a biophysical score \( \mathcal{B}(C_j) \) and ensure that it exceeds 0.55, which is the minimum acceptable score determined by the lowest biophysical score observed in natural proteins. The final selection probability is determined by:
\begin{equation}
    P_{\text{select}}(C_j) = \frac{\exp(\mathcal{B}(C_j) / \tau_t)}{\sum_{k=1}^K \exp(\mathcal{B}(C_k) / \tau_t)},
\end{equation}
where \( \tau \) is the adaptive diversity control parameter at the current step, controlled by the decay schedule:
\begin{equation}
\tau \leftarrow \max(\tau_{\text{final}}, \tau \cdot \gamma).
\end{equation}

This adaptive diversity control mechanism enables a dynamic exploration-exploitation trade-off throughout the generation process. During early stages (high \( \tau \)), the model maintains broad exploration capability by accepting a wider range of candidates, including those with suboptimal biophysical scores. As generation progresses and \( \tau \) decreases, the selection becomes increasingly focused on candidates with superior biophysical properties.

% \begin{figure}[h]
%     \centering
%     \includegraphics[width=0.7\linewidth]{fig/tau.png}
%     \caption{The decay schedule of the Adaptive Diversity Control Strategy. The initial value of \( \tau_t \) is 1.0, the final value is 0.001, and the decay rate is 0.1.}
%     \label{fig:adc}
% \end{figure}

This joint selection mechanism, enhanced by adaptive diversity control, ensures that the generated sequences are not only statistically coherent and biologically viable, but also maintain evolutionary diversity. The dynamic adjustment of \( \tau \) prevents premature convergence to local optima while gradually guiding the selection toward functionally promising regions of the sequence space.

\begin{figure*}[t]
  \centering
  \includegraphics[width=1\textwidth]{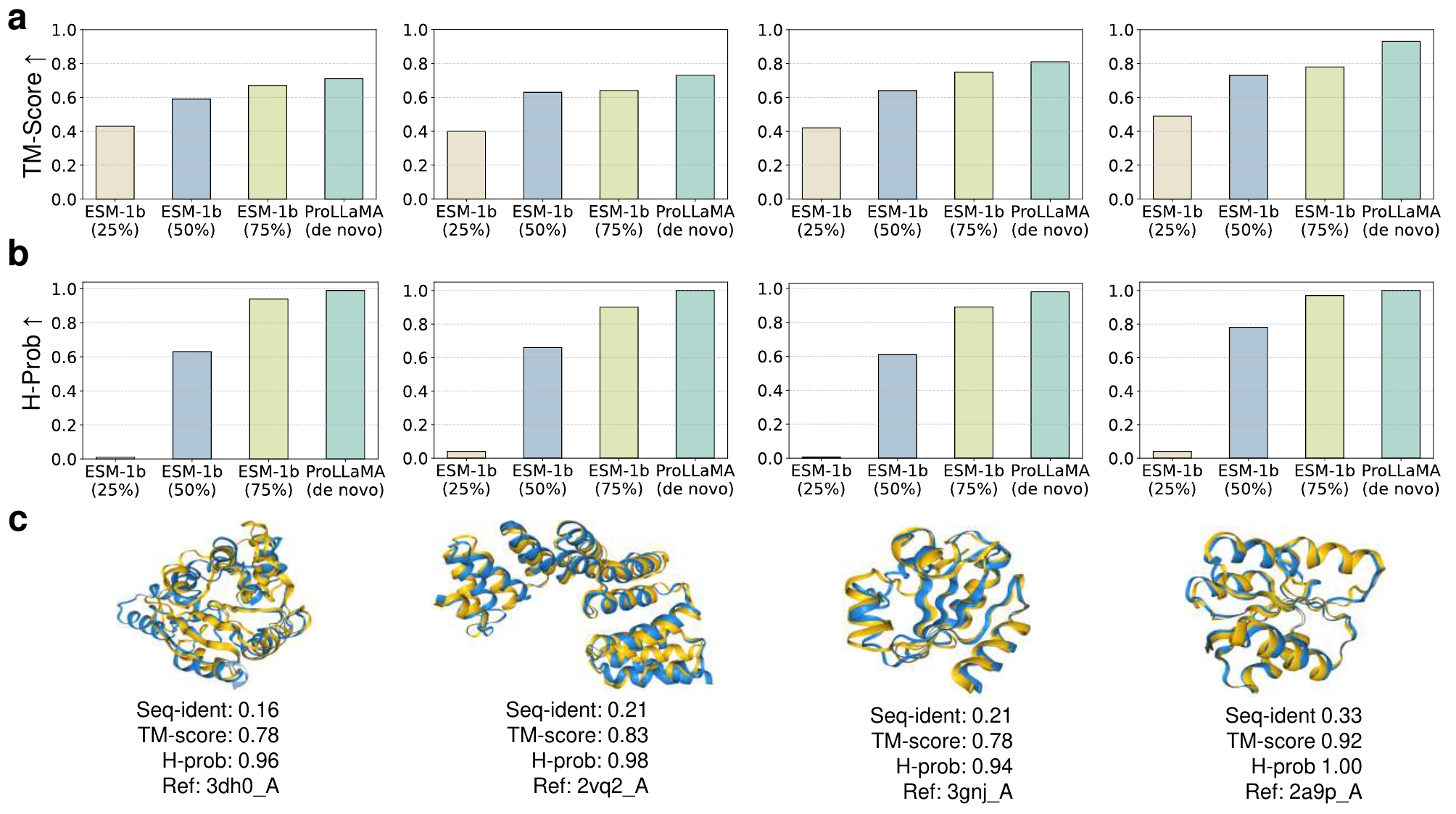}
  \vspace{-1.5\baselineskip}
  \caption{\textbf{The performance of ProLLaMA in Conditional Protein Generation.} Proteins of four superfamilies generated by our ProLLaMA outperform ESM-1b in terms of (a) TM-score and (b) H-Prob, even partial residues are provided to ESM-1b. $25\%$: 25\% residues are provided. (c) We visualize four proteins generated by ProLLaMA using SAM-MT, TPHD, Trx, and CheY as instructions. Blue is generated protein, and yellow is the natural protein as the reference, indicating that the generated protein is novel in sequence and reliable in structure.}
  \label{fig:esm_prollama}
\end{figure*}

\begin{figure*}[t]
  \centering
  \includegraphics[width=1\textwidth]{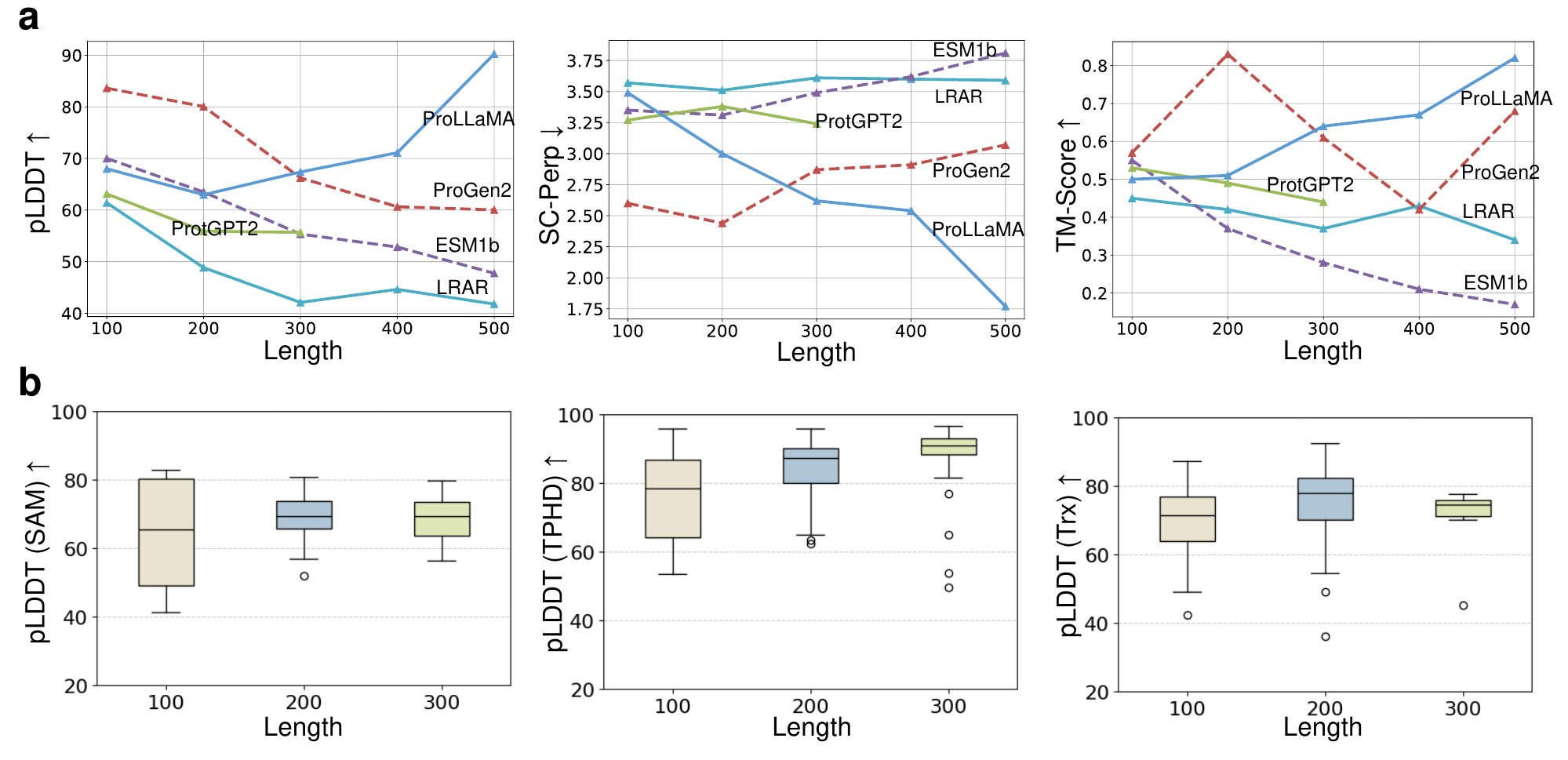}
  \vspace{-2\baselineskip}
  \caption{\textbf{Comparison of proteins across different length intervals.} (a) Unconditional Generation: Compared to other methods, ProLLaMA maintains a high quality of generated proteins as their length increases. (b) Conditional Generation: Distribution of pLDDT values of proteins generated by ProLLaMA in different length intervals, validating the effectiveness of ProLLaMA.}
  \label{fig:length_dist}
\end{figure*}

\section{Experiments} \label{sec:experiment}
We introduce the experiment setup in Section~\ref{sec:metrics}. And we evaluate the unconditional protein generation task in Section~\ref{sec:uncondition_sequence_generation}, 
the controllable protein generation task in Section~\ref{sec:controllable_generation}, the protein property prediction task in Section~\ref{sec:property_prediction}.

\subsection{Experiment Setup} \label{sec:metrics}
\textbf{Training Settings:} For continued pre-training, the LoRA rank is set to 128, employing the AdamW optimizer alongside a cosine annealing scheduler with a warm-up. The peak learning rate stands at 5e-5, with a total of one training epoch. It takes six days on eight A6000 GPUs using FlashAttention-2~\citep{dao2023flashattention}. For instruction tuning, the LoRA rank is set to 64 with two training epochs, and all other settings remain consistent with the continued pre-training setup. It takes 5 days on eight A6000 GPUs. More training details can be found in Supplementary Material.

\textbf{Evaluation Settings:} Unconditional protein generation involves generating protein sequences without specific instructions. Controllable protein generation involves generating desired protein sequences based on instructions that specify the required superfamily. Property prediction involves predicting protein superfamily and solubility based on instructions, which include the protein sequences to be predicted. All evaluations are conducted on one GPU with 24GB of VRAM. For EPGF, the number of candidates is 8, the initial $\tau$ is 1.0, the final $\tau$ is 1.0, and the decay rate is 0.1.

\textbf{Evaluation Metrics:} We use the following metrics to evaluate the generated protein sequences. The pLDDT~\citep{jumper2021highly} is used to measure whether the sequences are structurally plausible. Self-Consistency Perplexity~(SC-Perp)~\citep{alamdari2023protein} serves as an additional metric of plausible structures since pLDDT falls short in dealing with intrinsically disordered regions~(IDRs)~\citep{davey2019functional}. The TM-score~\citep{zhang2004scoring} reflects the structural similarity between the generated sequences and the known ones in AFDB~\citep{varadi2022alphafold} and PDB~\citep{berman2002protein}. RMSD also reflects the structural similarity from the perspective of atomic distance. 
Homologous probability~(H-Prob) reflects the probability that the generated protein is homologous to a known one.
Seq-Ident reflects the sequence similarity between generated sequences and known ones. More details are shown in Supplementary Material.

\textbf{Baselines:} The baselines cover various types of models. As shown in Table~\ref{table:prollama-compare}, CARP and LRAR belong to Convolutional Neural Networks~(CNN). ESM-1b and ESM-2 are language models based on Auto-Encoder~(AE) architectures, and we use Gibbs sampling to make them generate proteins. EvoDiff is a diffusion model. ProtGPT2, Mol-Instructions, and our ProLLaMA are Auto-Regressive~(AR) language models. We demonstrate the comparison of the parameters in Supplementary Material.

\subsection{Unconditional Protein Generation} \label{sec:uncondition_sequence_generation}

% Unconditional protein sequence generation can assess whether the model has learned the protein language well.
Table~\ref{table:prollama-compare} shows the results. Our ProLLaMA is optimal on pLDDT, TM-score, and RMSD and suboptimal on SC-Perp. This indicates that ProLLaMA, through its training on protein sequence data, can generate structurally plausible proteins.
%Furthermore, ProLLaMA does not just memorize the training dataset, it learns general rules, as the Seq-Ident value of 16.68 (below 20) indicates low sequence similarity. 
In particular, ProLLaMA-generated proteins exhibit a mean and standard deviation for pLDDT and SC-Perp of 66.49±12.61 and 3.10±0.65, respectively. These values are comparable to those of natural proteins as reported in~\citep{alamdari2023protein}, which are 68.25±17.85 and 3.09±0.63, respectively.

The de novo design of long and structurally plausible protein sequences is a huge challenge~\citep{ferruz2022protgpt2}, yet our ProLLaMA performs well.
As shown in Fig.~\ref{fig:length_dist}(a), when the length is greater than 300, ProLLaMA performs the best in all three metrics. Although ProGen2's performance is better in short sequences (length$\leq$200), it decreases as the length increases. This indicates that ProLLaMA is capable of capturing long-range dependencies between amino acids while other models struggle.

\subsection{Controllable Protein Generation} \label{sec:controllable_generation}
We use four superfamily descriptions as instructions respectively: the S-adenosyl-L-methionine-dependent methyltransferase superfamily (SAM-MT), the Tetratricopeptide-like helical domain superfamily (TPHD), the Thioredoxin-like superfamily (Trx), and the CheY-like superfamily (CheY). For each superfamily, ProLLaMA generates 100 protein sequences.
We randomly select 100 natural proteins from each of the four superfamilies as benchmarks for comparison. We employ Foldseek~\citep{van2024fast} to compare generated proteins with natural ones.
% We use four superfamilies for experiments: S-adenosyl-L-methionine-dependent methyltransferase superfamily~(SAM-MT), the Tetratricopeptide-like helical domain superfamily~(TPHD), the Thioredoxin-like superfamily~(Trx), and CheY-like superfamily~(CheY). For each of the four superfamilies, we make ProLLaMA generate 100 protein sequences. We also take 100 natural proteins belonging to the four superfamilies respectively as benchmarks and comparison objects. We use Foldseek to compare the generated proteins and the natural proteins. We first deduplicate the comparison results, and then calculate the average values of TM-score and H-Prob. 

\textbf{The protein generation of ProLLaMA is controllable.} The TM-scores shown in Table~\ref{table:prollama-conditional-compare} demonstrate that ProLLaMA can generate desired protein sequences based on instructions that specify the required functionalities, confirming the capability for controllable generation. For SAM-MT and Trx, the TM-scores of our generated sequences are about 0.8; for TPHD and CheY, they are around 0.9. The high TM-score indicates that the structures of the generated proteins closely resemble those of natural proteins in the same superfamily, implying functional similarity. In contrast, other models exhibit significantly lower TM-score due to their lack of controllable generation. 

Furthermore, ProLLaMA's de novo generation outperforms ESM-1b's non-de novo generation, even when ESM-1b is provided with 75\% of the residues, as evidenced by higher TM-scores and H-Prob in Fig.~\ref{fig:esm_prollama}(a)(b).  As shown in Fig.~\ref{fig:length_dist}(b), ProLLaMA demonstrates robust generation performance across different sequence lengths. Across all superfamilies, the proteins generated by ProLLaMA maintain a high pLDDT across different length ranges. The results in Supplementary Material also indicate that their TM-Score is high, around 0.9.
These results highlight ProLLaMA's ability to effectively capture structural and evolutionary relationships through text and sequence learning, enabling precise control over protein generation. 
% Additionally, although Mol-Instructions is capable of controllable protein generation in principle, its lower TM-score and pLDDT values indicate a limited understanding of structurally plausible protein sequences.

\textbf{EPGF benefits ProLLaMA to generate more biologically plausible proteins.} 
We compared the performance of ProLLaMA with and without EPGF across four aforementioned protein superfamilies. As shown in Fig.~\ref{fig:epgf}, ProLLaMA+EPGF consistently outperformed ProLLaMA alone. Specifically, ProLLaMA+EPGF achieved higher pLDDT scores, indicating greater confidence in local structure prediction, with mean values closer to or even exceeding those of natural proteins in the TPHD and CheY superfamilies (85.5 vs. 83.7, 84.4 vs. 83.9). Additionally, ProLLaMA+EPGF generated sequences with significantly higher BioScores, demonstrating better adherence to evolutionary and biophysical constraints. Finally, the TM-scores of ProLLaMA+EPGF were consistently higher, particularly in the TPHD and Trx superfamilies, indicating greater structural similarity to natural proteins. These results highlight EPGF's ability to guide ProLLaMA in generating biologically viable and structurally coherent protein sequences, bridging the gap between statistical language modeling and functional protein design. We provide more results and ablation experiments of EPGF in Supplementary Material.

\begin{table}[t]
\caption{\textbf{Controllable generation of ProLLaMA.} SAM-MT, TPHD, Trx, and CheY are four supefamiles.}
\begin{center}
\scalebox{1}{
\begin{tabular}{l|c|c|c|c}
    \toprule
    \textbf{Methods} & \textbf{SAM-MT}  & \textbf{TPHD} &  \textbf{Trx}  & \textbf{CheY}\\
    \midrule
     ESM-1b &  0.58 & 0.55 & 0.61 &  0.63 \\
     ESM-2 &  0.52  & 0.51   & 0.53  & 0.57  \\
     EvoDiff & 0.46 & 0.42 & 0.42  & 0.46 \\
     ProtGPT2 &0.45  & 0.43  & 0.44  & 0.45  \\
     ProGen2 &  0.44 & 0.45  &0.43 & 0.44  \\
     Mol-Instructions & 0.39 & 0.38 & 	0.39&  0.45\\
     % \textbf{ProLLaMA}   & \textbf{{0.71}}  & \textbf{{0.82}}& \textbf{{0.81}} & \textbf{{0.93}} \\
     \textbf{ProLLaMA}   & {0.85}  & {0.89}& {0.79} & {0.95} \\
     \textbf{ProLLaMA+EPGF}   & \textbf{0.85} & \textbf{0.95} & \textbf{0.86} & \textbf{0.96} \\
    \bottomrule
\end{tabular}
}
\end{center}
\label{table:prollama-conditional-compare}
\end{table}

\begin{table}[t]
    \centering 
    \caption{\textbf{Protein superfamily prediction.}}
    \scalebox{1}{
    \begin{tabular}{c|c|c} % 5 行 2 列的表格
        \toprule
        \textbf{Metric} & \textbf{5-fold Validation} &\textbf{Test} \\
        \midrule
        Accuracy &  0.671$\pm$0.005  & 0.671 \\
        Precision & 0.702$\pm$0.004  & 0.701 \\
        Recall &    0.700$\pm$0.005    &0.697 \\
        Jaccard &      0.691$\pm$0.004  &0.690 \\
        \bottomrule
    \end{tabular}
    }
   \label{table:whole_test_set}\vspace{-1\baselineskip}
\end{table}

\begin{table}[t]
\caption{\textbf{Protein Solubility Prediction.} *:Values are sourced from GraphSol~\citep{chen2021structure}.} \label{table:appendix_solubility}
\small
\centering
\scalebox{0.9}{
\begin{tabular}{lccccc}
\toprule
\textbf{Method}& \textbf{Accuracy} &\textbf{Precision} & \textbf{Recall} & \textbf{F1-score} \\%& \textbf{AUC} \\
\midrule
Protein-Sol*~\citep{hebditch2017protein}& 0.714&	0.689&  0.688&	0.693	\\%& 0.808 \\
DeepSol*~\citep{khurana2018deepsol}& 0.763 &	 0.771     &	\textbf{0.738} &	 0.695	\\%& 0.845 \\
GraphSol*~\citep{chen2021structure}& \textbf{0.779} &	 0.775     &	0.693 &	 0.732	\\%& 0.866 \\
ProLLaMA (ours)& 0.775 &	 \textbf{0.788}     &	0.685 &	 \textbf{0.733}	\\%& 0.767 \\
\bottomrule
\end{tabular}
}
\end{table}

\textbf{Case study of four generated proteins.} In Fig.~\ref{fig:esm_prollama}(c), we visualize four examples of proteins generated by ProLLaMA (colored in blue) alongside the most structurally similar natural proteins from PDB as reference(colored in yellow). The significant overlap in 3D structures and the high TM-score confirm structural similarity. Low Seq-ident indicates sequence diversity. In summary, through controllable protein generation, ProLLaMA is capable of generating desired proteins with structures similar to natural proteins, yet with novel sequences.

% Overall, ProLLaMA can generate protein sequences that are structurally similar to natural proteins but with distinct sequences through controlled generation.

\subsection{Property Prediction} \label{sec:property_prediction}

\textbf{Superfamily Prediction}. We use the test dataset to evaluate whether ProLLaMA can predict the superfamily to which a given protein belongs. The test dataset consists of 10,000 samples.
Although ProLLaMA performs a classification task here, it is more complex than typical ones.
The key difference is that typical classification tasks require models to output a fixed label, often in one-hot encoding. 
In contrast, ProLLaMA outputs the text. 
The advantage of the latter lies in its flexibility, such as the ability to easily handle situations where a sample belongs to multiple categories simultaneously. However, this increases task difficulty due to the much larger number of potential classification categories.

As shown in Table~\ref{table:whole_test_set}, our model achieves an accuracy of 67.1\% in predicting protein superfamilies on the test set, matching the performance observed in 5-fold validation (67.1\% ± 0.5\%). The model achieves a precision of 70.1\%, recall of 69.7\%, and a Jaccard score of 69.0\% on the test set, all closely aligned with the validation results These results indicate the model's robustness and generalizability. The calculation formulas for these metrics can be found in Supplementary Material.

\textbf{Solubility Prediction.} We transform the eSol dataset~\cite {niwa2009bimodal,hanson2018accurate} into an additional instruction dataset, which includes two tasks: generating proteins based on solubility and determining the solubility of proteins. We binarize the solubility, with \enquote{Solubility is False} indicating insoluble and \enquote{Solubility is True} indicating soluble.

We train ProLLaMA (the one after the first training stage) on this additional instruction dataset. The LoRA rank is 64, the learning rate is 5e-5, and the number of training steps is 370. We compare our ProLLaMA with other methods in predicting protein solubility. The results are shown in Table.~\ref{table:appendix_solubility}. Our ProLLaMA outperforms models specifically designed for solubility prediction in terms of precision and F1-score. And its accuracy is almost the same as that of GraphSol. In particular, ProLLaMA utilizes only protein sequences, whereas other models incorporate additional features such as absolute charge, secondary structure probabilities, etc (See Supplementary Material for detailed discussion).

\section{Related Work}

\textbf{Protein Language Models.} Recognizing the similarity between natural language sequences and protein sequences, many methods of NLP have been applied to protein sequence data~\citep{yang2018learned,alley2019unified,rao2021msa,elnaggar2021prottrans}. This has led to the development of PLMs, which are broadly categorized into two types~\citep{ferruz2022controllable,zheng2023structure}: Auto-Regressive~(AR) PLMs and Auto-Encoder~(AE) PLMs. AR PLMs adopt decoder-only architecture and Causal Language Modeling~(CLM)~\citep{bengio2000neural,vaswani2017attention}. They mainly concentrate on PLG~\citep{moffat2022design,ferruz2022protgpt2,madani2023large,nijkamp2023progen2}, with a minority also focusing on fitness prediction~\citep{pmlr-v162-notin22a}. 
AE PLMs adopt the encoder-only architecture and Masked Language Modeling~(MLM)~\citep{devlin2018bert,meier2021language,rives2021biological,brandes2022proteinbert,lin2023evolutionary}. They excel in PLU, with the learned protein representations applied to downstream predictive tasks~\citep{xu2023protst}. However, they face challenges in de novo protein generation. Our ProLLaMA is capable of multitasking, excelling in tasks in which both types specialize, and surpassing existing PLMs. This multitasking capability is achieved through instruction following, making it user-friendly. We have also noticed the recent emergence of scientific LLMs~\citep{fang2023mol,pei2024biot5+,yang2024zhongjing,taylor2022galactica,pei2024leveraging}.
Although these models can also address certain protein-related problems, they lack a deep understanding of protein sequences due to the absence of large-scale pretraining on protein-specific corpora. As a result, they are better suited for handling general scientific question-answering tasks, but usually perform poorly when faced with complex protein tasks, especially protein generation.

\section{Conclusion}
Existing PLMs excel in either protein generation tasks or protein understanding tasks.
In this work, we introduce an efficient training framework to transform any general LLM into a multi-task PLM. We construct an instruction dataset containing both generation tasks and understanding tasks. We developed ProLLaMA, a versatile PLM for multiple tasks such as controllable protein generation and prediction of protein properties. Evolutionary Protein Generation Framework~(EPGF) plays a crucial role in bridging the gap between statistical language modeling and biological constraints.
Experiments indicate that ProLLaMA and EPGF perform exceptionally well.
We are confident that our work will have a significant impact on the AI4Science community and also open up exciting avenues for further exploration in biologically grounded generation strategies such as optimized EPGF.

\bibliographystyle{unsrt}
\bibliography{refer}{}

\end{document}